\documentclass{IEEEcsmag}
\usepackage[utf8]{inputenc}
\usepackage{color}
\usepackage{epsfig}

\usepackage{inconsolata}
\usepackage{tikz}
\usetikzlibrary{arrows,shapes,shadows}
\definecolor{guixorange1}{RGB}{243,154,38}  
\definecolor{guixblue2}{RGB}{10,50,80} 
\definecolor{guixred2}{RGB}{230,68,57}  
\definecolor{guixdarkgrey}{RGB}{46,47,55} 
\DeclareUnicodeCharacter{2026}{\textrm{\ldots}}
\definecolor{darkblue}{rgb}{0.0, 0.0, 0.55}
\definecolor{cobalt}{rgb}{0.0, 0.28, 0.67}
\definecolor{coolblack}{rgb}{0.0, 0.18, 0.39}
\PassOptionsToPackage{hyphens}{url}\usepackage[colorlinks=true, linkcolor=coolblack, urlcolor=cobalt, citecolor=coolblack]{hyperref}
\RequirePackage[hyphens]{url}
\newdimen\oldframetabcolsep
\newdimen\oldcolortabcolsep
\newdimen\oldpretabcolsep

\title{Reproducibility and Performance{\char58} Why Choose?}
\author{{Ludovic Court\`{e}s}}
\affil{
Inria}

\begin{document}

\begin{abstract}
Research processes often rely on high-performance %
computing (HPC), but HPC is often seen as antithetical to %
“reproducibility”{\char58} one would have to choose between software that %
achieves high performance, and software that can be deployed in a %
reproducible fashion.  However, by giving up on reproducibility we would %
give up on verifiability, a foundation of the scientific process.  How %
can we conciliate performance and reproducibility?  This article looks %
at two performance-critical aspects in HPC{\char58} message passing (MPI) and %
CPU micro-architecture tuning.  Engineering work that has gone into %
performance portability has already proved fruitful, but some areas %
remain unaddressed when it comes to CPU tuning.  We propose package %
multi-versioning, a technique developed for GNU Guix, a tool for %
reproducible software deployment, and show that it allows us to %
implement CPU tuning without compromising on reproducibility and %
provenance tracking.
\end{abstract}

\maketitle

\chapterinitial{Introduction.}
It should come as no surprise that the execution speed of programs is a %
primary concern in high-performance computing (HPC).  Many HPC %
practitioners would tell you that, among their top concerns, is the %
performance of high-speed networks used by the Message Passing Interface %
(MPI) and use of the latest vectorization extensions of modern CPUs.\par
This article focuses on the latter{\char58} tuning code for specific CPU %
micro-architectures, to reap the benefits of modern CPUs.  This question %
is particularly acute in the context of GNU Guix, a software deployment %
tool with strong support for {\em{reproducible deployment}}.  We like %
to present Guix as a key element of the reproducible research toolbox{\char58} %
as more research output is produced by software, the ability to {\em{verify and validate}} research results depends on the ability to {\em{re-deploy and re-run}} the software.  We present a recently-introduced %
CPU-tuning option for Guix, the design choices we made, and how this %
affects reproducibility.\par
But let us first consider this central question in the HPC and %
scientific community{\char58} can “reproducibility” be achieved {\em{without}} sacrificing performance?  Our answer is a resounding “yes”, %
but that deserves clarifications.\par

\section*{Reproducibility \& High Performance}
The author remembers advice heard at the beginning of their %
career in HPC—advice still given today—{\char58} that to get optimal MPI %
performance, you would have to use the vendor-provided MPI library; that %
to get your code to perform well on this new cluster, you would have to %
recompile the complete software stack locally; that using generic, %
pre-built binaries from a GNU/Linux distribution will not give you good %
performance.\par
From a software engineering viewpoint, this looks like a sad %
situation and an inefficient approach, dismissing the benefits of %
automated software deployment as pioneered by Debian, Red Hat, and %
others in the 90’s or, more recently, as popularized with container %
images.  It also means doing away with reproducibility, where %
“reproducibility” is to be understood in two different ways{\char58} first as %
the ability to re-deploy the same software stack on another machine or %
at a different point in time, and second as the ability to {\em{verify}} that %
binaries being run match the source code—the latter is what reproducible %
builds are concerned with [10].\par
But does it really have to be this way?  Engineering efforts to %
support {\em{performance portability}} suggest otherwise.  A mature %
MPI implementation like Open MPI, today, does achieve performance %
portability{\char58} it takes advantage of high-speed networking hardware by %
determining, at run-time, which drivers to use to obtain optimal %
performance for the network at hand—no recompilation is needed [4].\par
Likewise, generic, pre-built binaries can and indeed often do %
take advantage of modern CPUs by selecting at run-time the most %
efficient implementation of performance-sensitive routines for the host %
CPU [3].  There are cases, though, where %
this is {\em{not}} the case; these are those we will focus on in the %
remainder of this article.\par

\section*{The Jungle of SIMD Extensions}
While major CPU architectures such as x86\_64, AArch64, and %
POWER9 were defined years ago, CPU vendors regularly extend them. %
Extensions that matter most in HPC are vector extensions{\char58} single %
instruction/multiple data (SIMD) instructions and registers.  In this %
area, a {\em{lot}} has happened on x86\_64 CPUs since the baseline %
instruction set architecture (ISA) was defined.  As shown in Figure~1, Intel and AMD have been %
tacking ever more powerful SIMD extensions to their CPUs over the years, %
from SSE3 to AVX-512, leading to a wealth of CPU “micro-architectures”. %
This gives a high-level view, but just looking at generations of Intel %
processors by their code name—from “Nehalem” to “Skylake” {\textit{via}} %
“Ivybridge”—shows an already more complicated story.\par
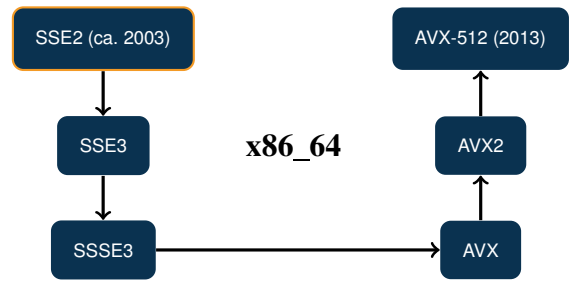
\begin{figure}[ht]
\begin{scriptsize}
  \begin{tikzpicture}[
        box/.style = { fill=guixblue2, text=white, inner sep=3mm, rounded corners, font=\bf\sf }
      ]
    \matrix[row sep=6mm, column sep=6mm] {
      \node(sse2) [box, draw=guixorange1, thick] {SSE2 (ca. 2003)}; & & \node(avx512) [box] {AVX-512 (2013)}; \\
      \node(sse3) [box] {SSE3}; & \node {\large\textbf{x86\_64}}; & \node(avx2) [box] {AVX2}; \\
      \node(ssse3) [box] {SSSE3}; & & \node(avx) [box] {AVX}; \\
    };

    \path[very thick, draw=guixorange1] (sse2) edge [->] (sse3);
    \path[very thick, draw=guixorange1] (sse3) edge [->] (ssse3);
    \path[very thick, draw=guixorange1] (ssse3) edge [->] (avx);
    \path[very thick, draw=guixorange1] (avx) edge [->] (avx2);
    \path[very thick, draw=guixorange1] (avx2) edge [->] (avx512);
  \end{tikzpicture}

\end{scriptsize}
\caption{\label{fig-simd-extensions}Timeline of x86\_64 SIMD extensions}\end{figure}
Linear algebra routines that scientific software relies on %
greatly benefit from SIMD extensions.  For example, on a modest Intel %
CORE i7 processor (of the Skylake generation), the AVX2-optimized %
version of the dense matrix multiplication routines of Eigen (\href{https://eigen.tuxfamily.org}{\textit{https{\char58}//eigen.tuxfamily.org}}), built with GCC 10.3, peaks at about 40 Gflops/s, %
compared to 11 Gflops/s for its baseline x86\_64 version—four times %
faster!\par

\section*{Portable Performance Through Function Multi-Versioning}
How to create binaries that are portable, yet are able to get %
the most out of the CPU on which they are executed?  This has been an %
important question for distributors of binaries.  Distributions such as %
Debian and CentOS provide the convenience of fast automated deployment, %
thanks to pre-built binaries; asking users to either recompile part of %
their software stack or give up on performance is not a reasonable %
alternative.\par
To address this and achieve performance portability, %
developers have largely adopted {\em{function multi-versioning}} %
(FMV){\char58} the implementation provides multiple versions of “hot” routines, %
one for each relevant CPU micro-architecture, and picks the best one for %
the host CPU at run time.  Many pieces of performance-critical software %
already use this technique{\char58} the C standard library (libc) contains %
multiple versions of its string handling and math routines, the GMP %
library for multi-precision arithmetic uses FMV, and so do software %
packages ranging from cryptography libraries (Libgcrypt, Nettle) to %
linear algebra (OpenBLAS, FFTW).\par
To make it easier for developers to adopt FMV, the GNU %
compilation tool chain (GCC, the Binary Utilities, and the C Library), %
which is widely used in HPC, provides helpers at different levels. %
Developers can annotate relevant functions with the {\texttt{target\_clone}} %
attribute to instruct the compiler to generate optimized versions of the %
function for each selected architecture.  GCC not only generates these %
versions, but also generates code to choose the right function version %
for the host CPU at load time, with support from the dynamic linker, %
{\texttt{ld.\-so}}.  That relieves developers from the need to implement %
their own ad-hoc machinery.  From that perspective, it would seem that %
performance portability, {\textit{via}} FMV, is a solved problem.\par
There is at least one common pattern though where FMV is not %
applicable, or at least is not applied{\char58} C++ header-only libraries. %
These are libraries that provide generic template code in header files; %
that code is specialized {\em{at build time}} in software that uses %
them.  There is no shortage of C++ header-only math libraries providing %
efficient, optimized SIMD versions of their routines{\char58} Eigen, MIPP, xsimd %
and xtensor, SIMD Everywhere (SIMDe), Highway, and many more.  All %
these, except Highway, have in common that they do {\em{not}} support %
FMV.  Since they “just” provide headers, it is up to {\em{each}} %
package using them to figure out what to do in terms of performance %
portability.\par
In practice though, software using these C++ header-only %
libraries rarely makes provisions for performance portability.  Thus, %
when compiling those packages for the baseline ISA, one misses out on %
all the vectorized implementations that libraries like Eigen provide. %
This is a known issue in search of a solution—see \href{https://gitlab.com/libeigen/eigen/-/issues/2344}{\textit{https{\char58}//gitlab.com/libeigen/eigen/-/issues/2344}}. %
It can have a very concrete impact on %
performance since many scientific packages—the ARPACK-NG library for %
solving eigenvalue problems, the Ceres solver for optimization problems, %
the FEniCSx platform for solving differential equations, to name a %
few—depend on Eigen.\par

\section*{Reproducible Deployment}
Distributions such as Debian and Fedora that provide pre-built %
binaries miss out on SIMD optimizations of C++ header-only libraries %
like Eigen because they provide binaries targeting the baseline CPU %
architecture so that those binaries run on any CPU.  The Spack [7] and EasyBuild [8] %
package managers address that by {\em{rebuilding}} software on the %
target computer, which allows them to instruct the compiler to optimize %
for the host CPU.\par
Unfortunately, EasyBuild and Spack both have limited support %
for reproducible deployment—they do not, in general, guarantee that you %
can redeploy the same software environment on different machines, or at %
different points in time.  This is because they build upon software %
provided by the host system—the compiler tool chain, “system” libraries, %
etc.—and that foundation differs from one system to another—e.g., CentOS %
might provide some version of GCC, and Ubuntu might provide another.\par
To avoid that, Guix builds software in {\em{isolated %
environments}}, as pioneered by Nix [1, 6], and its package collection is {\em{self-contained}}—it does not rely on external software packages.  This %
is what makes Guix builds reproducible bit-for-bit—or in other words, %
{\em{verifiable}} [10].  Given binaries %
and provenance data, anyone can independently verify the %
binary/source-code correspondence.\par
Guix provides a command-line interface similar to that of %
other package managers{\char58} {\texttt{guix install pyth\-on}}, for instance, %
installs the Python interpreter.  Package management is per-user rather %
than system-wide and does not require system administrator privileges, %
which makes it suitable for multi-user HPC clusters [2].  To offer the level of flexibility that HPC %
users expect, Guix lets users customize packages {\textit{via}} {\em{package transformation options}} on the command line—for instance to %
swap two packages in the dependency graph—or through programming %
interfaces [2].\par
Quite uniquely, Guix supports {\em{“time traveling”}}{\char58} with %
{\texttt{guix time-\-mach\-ine}}, users can run a specific revision of Guix and %
use it to deploy packages as they were defined in that revision.  The %
typical use case is redeploying software that was used to produce %
computational results for a scientific publication [5, 9, 11].  The command below deploys Python, NumPy, and %
their dependencies as they were defined in a Guix revision from October %
2021{\char58}\par

\vspace{3mm}
\begin{footnotesize}
\setlength{\oldpretabcolsep}{\tabcolsep}
\addtolength{\tabcolsep}{-\tabcolsep}
{\setbox1 \vbox \bgroup
{\noindent \texttt{\begin{tabular}{l}
guix\ time-machine\ --commit=b0735c79b0d1d341\ --\ {\char92}\\
\ \ shell\ python\ python-numpy\\
\\
\end{tabular}
}}
\egroup{\box1}}%
\setlength{\tabcolsep}{\oldpretabcolsep}

\end{footnotesize}
Whether you run it today or two years from now, it will deploy %
the {\em{exact same binaries}}, bit-for-bit, down to the C %
library.\par

\section*{Package Multi-Versioning}
With our packaging hammer, one could envision a solution to %
these CPU tuning problems{\char58} if we cannot do function multi-versioning, %
what about implementing {\em{package}} multi-versioning?  Guix makes %
it easy to define package variants, so we can define package variants %
optimized for a specific CPU—compiled with {\texttt{-\-march\-=skylake}}, for %
instance.  What we need is to define those variants “on the fly”.\par
The recently-introduced {\texttt{-\--\-tune}} package transformation %
option works along those lines.  Users can pass {\texttt{-\--\-tune}} to any of %
the command-line tools ({\texttt{guix install}}, {\texttt{guix sh\-ell}}, etc.) %
and that causes “tunable” packages to be optimized for the host CPU. %
For example, here is how you would run Eigen’s matrix multiplication %
benchmark from the {\texttt{eigen-\-bench\-marks}} package with %
micro-architecture tuning{\char58}\par

\vspace{3mm}
\begin{footnotesize}
\setlength{\oldpretabcolsep}{\tabcolsep}
\addtolength{\tabcolsep}{-\tabcolsep}
{\setbox1 \vbox \bgroup
{\noindent \texttt{\begin{tabular}{l}
\$\ guix\ shell\ --tune\ eigen-benchmarks\ --\ {\char92}\\
\ \ \ \ benchBlasGemm\ 240\ 240\ 240\\
guix\ shell{\char58}\ tuning\ for\ CPU\ skylake\\
240\ x\ 240\ x\ 240\\
cblas{\char58}\ 0.208547\ (15.908\ GFlops/s)\\
eigen\ {\char58}\ 0.0720303\ (46.06\ GFlops/s)\\
l1{\char58}\ 32768\\
l2{\char58}\ 262144\\
\\
\end{tabular}
}}
\egroup{\box1}}%
\setlength{\tabcolsep}{\oldpretabcolsep}

\end{footnotesize}
{\texttt{-\--\-tune}} determines the name of the host CPU as %
recognized by GCC’s (and Clang’s) {\texttt{-\-march\-}} option.  Users can %
override auto-detection by passing a CPU name—e.g., {\texttt{-\--\-tune=skylake-\-avx512}}.  As mentioned earlier, we made the conscious %
choice of letting {\texttt{-\--\-tune}} affect solely software that packagers %
explicitly marked as “tunable”.  This ensures Guix does not end up %
rebuilding packages that could not possibly benefit from %
micro-architecture-specific optimizations, which would be a waste of %
resources.\par
This implementation of package multi-versioning does not %
sacrifice reproducibility.  When {\texttt{-\--\-tune}} is used, from Guix’s %
viewpoint, it is just an alternate, but well-defined dependency graph %
that gets built.  Guix records package transformation options that were %
used so it can “replay” them.  For example, one can export a {\em{manifest}} representing packages that have been deployed{\char58}\par

\vspace{3mm}
\begin{footnotesize}
\setlength{\oldpretabcolsep}{\tabcolsep}
\addtolength{\tabcolsep}{-\tabcolsep}
{\setbox1 \vbox \bgroup
{\noindent \texttt{\begin{tabular}{l}
\$\ guix\ shell\ eigen-benchmarks\ --tune\\
guix\ shell{\char58}\ tuning\ for\ CPU\ skylake\\
{\char91}env{\char93}\$\ guix\ package\ --export-manifest\ {\char92}\\
\ \ \ \ \ \ \ \ \ \ \ \ \ \ -p\ \$GUIX\_ENVIRONMENT\\
(use-modules\ (guix\ transformations))\\
\\
(define\ transform1\\
\ \ (options-$>$transformation\\
\ \ \ \ '((tune\ .\ "skylake"))))\\
\\
(packages-$>$manifest\\
\ \ (list\ (transform1\\
\ \ \ \ \ \ \ \ \ \ (specification-$>$package\\
\ \ \ \ \ \ \ \ \ \ \ \ "eigen-benchmarks"))))\\
\\
\end{tabular}
}}
\egroup{\box1}}%
\setlength{\tabcolsep}{\oldpretabcolsep}

\end{footnotesize}
The manifest above is a code snippet that can be passed to %
{\texttt{guix sh\-ell}} or {\texttt{guix package}} to redeploy the package with %
the same tuning parameters.  Like other transformation options, {\texttt{-\--\-tune}} is accepted by all the commands; for example, here is how you %
would build a Docker image tuned for a particular CPU{\char58}\par

\vspace{3mm}
\begin{footnotesize}
\setlength{\oldpretabcolsep}{\tabcolsep}
\addtolength{\tabcolsep}{-\tabcolsep}
{\setbox1 \vbox \bgroup
{\noindent \texttt{\begin{tabular}{l}
guix\ pack\ -f\ docker\ -S\ /bin=bin\ \\
\ \ eigen-benchmarks\ --tune=skylake\\
\\
\end{tabular}
}}
\egroup{\box1}}%
\setlength{\tabcolsep}{\oldpretabcolsep}

\end{footnotesize}

\section*{Conclusion and Outlook}
We implemented what we call “package multi-versioning” for %
C/C++ software that lacks function multi-versioning and run-time %
dispatch, a notable example of which is optimized C++ header-only %
libraries.  It is another way to ensure that users do not have to trade %
reproducibility for performance.\par
The scientific programming landscape has been evolving over %
the last few years.  It is encouraging to see that Julia offers function %
multi-versioning for its “system image”, and that, similarly, Rust %
supports it with annotations similar to GCC’s {\texttt{target\_clones}}. %
Hopefully these new development environments will support performance %
portability well enough that users and packagers will not need to worry %
about it.\par
But first and foremost, it is up to us, research software %
engineers and scientists, to dispel the myth that performance is a valid %
excuse for non-reproducible computational workflows.\par

\section*{References}
\begin{flushleft}{%
\sloppy
\sfcode`\.=1000\relax
\newdimen\bibindent
\bibindent=0em
\begin{list}{}{%
        \settowidth\labelwidth{[21]}%
        \leftmargin\labelwidth
        \advance\leftmargin\labelsep
        \advance\leftmargin\bibindent
        \itemindent -\bibindent
        \listparindent \itemindent
        \itemsep 0pt
    }%
\item[{\char91}1{\char93}] L. Court\`{e}s. Functional Package Management with Guix. In {\textit{European Lisp Symposium}}, June 2013.
\item[{\char91}2{\char93}] L. Court\`{e}s,  R. Wurmus. Reproducible and User-Controlled Software Environments in HPC with Guix. In {\textit{Euro-Par 2015{\char58} Parallel Processing Workshops}}, Lecture Notes in Computer Science, pp. 579--591, August 2015.
\item[{\char91}3{\char93}] L. Court\`{e}s. Pre-Built Binaries vs. Performance. January 2018. {\textit{https{\char58}//hpc.guix.info/blog/2018/01/pre-built-binaries-vs-performance/}}.
\item[{\char91}4{\char93}] L. Court\`{e}s. Optimized and Portable Open MPI Packaging. December 2019. {\textit{https{\char58}//hpc.guix.info/blog/2019/12/optimized-and-portable-open-mpi-packaging/}}.
\item[{\char91}5{\char93}] L. Court\`{e}s. {\char91}Re{\char93} Storage Tradeoffs in a Collaborative Backup Service for Mobile Devices. In {\textit{ReScience C}}, 6(1) , June 2020, .
\item[{\char91}6{\char93}] E. Dolstra,  M. d. Jonge,  E. Visser. Nix{\char58} A Safe and Policy-Free System for Software Deployment. In {\textit{Proceedings of the 18th Large Installation System Administration Conference (LISA '04)}}, pp. 79--92, USENIX, November 2004.
\item[{\char91}7{\char93}] T. Gamblin,  M. LeGendre,  M. R. Collette,  G. L. Lee,  A. Moody,  B. R. d. Supinski,  S. Futral. The Spack Package Manager{\char58} Bringing Order to HPC Software Chaos. In {\textit{Proceedings of the International Conference for High Performance Computing, Networking, Storage and Analysis}}, SC '15, Association for Computing Machinery, 2015.
\item[{\char91}8{\char93}] M. Geimer,  K. Hoste,  R. McLay. Modern Scientific Software Management Using EasyBuild and %
Lmod. In {\textit{Proceedings of the First Workshop on HPC User Support %
Tools (HUST`14)}}, pp. 41--51, IEEE Press, 2014.
\item[{\char91}9{\char93}] K. Hinsen. Staged Computation{\char58} The Technique You Did Not Know You Were Using. In {\textit{Computing in Science Engineering}}, 22(4) ,  2020, pp. 99--103.
\item[{\char91}10{\char93}] C. Lamb,  S. Zacchiroli. Reproducible Builds{\char58} Increasing the Integrity of Software Supply Chains. In {\textit{IEEE Software}}, 39(2) , March 2022, pp. 62–70.
\item[{\char91}11{\char93}] J. M. Perkel. Challenge to Scientists{\char58} Does Your Ten-Year-Old Code Still Run?. In {\textit{Nature}}, 584, August 2020, pp. 656–658.

\end{list}}
\end{flushleft}

\begin{IEEEbiography}{Ludovic Courtès}
is a research software engineer at Inria, France.  He has been %
contributing to the development of GNU Guix since its inception in 2012 %
and works on its use in support of reproducible research workflows.  He %
holds a PhD in computer science from LAAS-CNRS.  You can reach him at %
{\textit{ludovic.courtes@inria.fr}}.
\end{IEEEbiography}

\end{document}